\documentclass[twocolumn]{aastex63}

\usepackage{amsmath}
\usepackage{natbib}
\usepackage{multirow} 
\usepackage{gensymb}
\usepackage[normalem]{ulem}
\useunder{\uline}{\ul}{}
\usepackage{hyperref}

\hypersetup{linkcolor=cyan,citecolor=magenta,filecolor=yellow,urlcolor=blue}

\hypersetup{linkcolor=magenta, citecolor=cyan, filecolor=yellow, urlcolor=blue}

\revised{\today}

\shorttitle{GRB 190114C: Fireball Energy Budget and Radiative Efficiency Revisited}
\shortauthors{Li \& Wang}

\begin{document}

\title{GRB 190114C: Fireball Energy Budget and Radiative Efficiency Revisited}

\author{Liang Li}
\affiliation{Institute of Fundamental Physics and Quantum Technology, Ningbo University, Ningbo, Zhejiang 315211, People's Republic of China}
\affiliation{School of Physical Science and Technology, Ningbo University, Ningbo, Zhejiang 315211, People's Republic of China}
\affiliation{INAF -- Osservatorio Astronomico d'Abruzzo, Via M. Maggini snc, I-64100, Teramo, Italy}

\author{Yu Wang}
\affiliation{ICRANet, Piazza della Repubblica 10, I-65122 Pescara, Italy}
\affiliation{ICRA and Dipartimento di Fisica, Universit\`a  di Roma ``La Sapienza'', Piazzale Aldo Moro 5, I-00185 Roma, Italy}
\affiliation{INAF -- Osservatorio Astronomico d'Abruzzo, Via M. Maggini snc, I-64100, Teramo, Italy}

\correspondingauthor{Liang Li, Yu Wang}
\email{liliang@nbu.edu.cn; yu.wang@uniroma1.it}

\begin{abstract}

The jet composition of gamma-ray bursts (GRBs), as well as how efficiently the jet converts its energy to radiation, are long-standing problems in GRB physics. Here, we reported a comprehensive temporal and spectral analysis of the TeV-emitting bright GRB 190114C. Its high fluence ($\sim$ 4.4$\times$10$^{-4}$ erg cm$^{-2}$) allows us to conduct the time-resolved spectral analysis in great detail and study their variations down to a very short time-scale ($\sim$0.1 s) while preserving a high significance. Its prompt emission consists of three well-separated pulses. The first two main pulses ($P_1$ and $P_2$) exhibit independently strong thermal components, starting from the third pulse ($P_3$) and extending to the entire afterglow, the spectra are all nonthermal, the synchrotron plus Compton upscattering model well interprets the observation. By combining the thermal ($P_1$ and $P_2$) and the nonthermal ($P_3$) observations based on two different scenarios (global and pulse properties) and following the method described in Zhang et al., we measure the fireball parameters and GRB radiative efficiency with little uncertainties for this GRB. A relevantly high GRB radiative efficiency is obtained based on both the global and pulse properties, suggesting that if GRBs are powered by fireballs, the efficiency can sometimes be high. More interestingly, though the observed parameters are individually different (e.g., the amount of mass loading $M$), the radiative efficiency obtained from $P_1$ ($\eta_\gamma=36.0\pm6.5\%$) and $P_2$ ($\eta_\gamma=41.1\pm1.9\%$) is roughly the same, which implies that the central engine of the same GRB has some common properties.

\end{abstract}

\keywords{Astronomy data analysis (1858); Gamma-ray bursts (629); Time domain astronomy (2109)}

\section{Introduction} \label{sec:intro}

On 2019 January 14 at 20:57:02.63 UT (hereafter $T_{0}$), an ultra-bright burst, GRB 190114C, was first detected by the Gamma-ray Burst Monitor (GBM) on board the NASA {\it Fermi} Gamma-ray Space Telescope \citep{Hamburg2019} and the Neil Gehrels {\it Swift} Observatory's Burst Alert Telescope \citep{J.D.Gropp2019}, and soon after Konus-\emph{Wind}, \emph{AGILE}/MCAL, \emph{INTEGRAL}/SPI-ACS, and the \emph{Insight-HXMT}/HE were triggered as well. Most interestingly, for the first time, TeV emission was detected by the Major Atmospheric Gamma Imaging Cherenkov (MAGIC) telescopes from $T_{0}$+57~s to $T_{0}$+15912 s, and its rich multi-wavelength observations were simultaneously observed from optical \citep{Bolmer2019,Castro-Tirado2019,Alexander2019,Tremou2019} to TeV \citep{MAGICCollaboration2019a} gamma-ray emissions. The duration ($T_{90}$), the time taken to accumulate $90\%$ of the burst fluence starting at the 5\% fluence level, reported by the Fermi/GBM Science Term is $\sim$116~s, and therefore, belong to the long-duration burst class. The 1024 ms peak flux and the fluence during the $T_{90}$ duration at 10-1000 keV measured by GBM are 246.864$\pm$0.859 photon cm$^{-2}$ s$^{-1}$ and (4.436$\pm$0.005)$\times$10$^{-4}$ erg cm$^{-2}$, respectively\footnote{\url{https://heasarc.gsfc.nasa.gov/W3Browse/fermi/fermigbrst.html}}. The measurement of redshift, $z$=0.424, has been reported by \cite{Castro-Tirado2019}. The isotropic energy with a $k$-correction to the rest-frame (1-10$^{4}$ keV), therefore, is estimated, $E_{\gamma,\rm iso}$=(2.48$\pm$0.22)$\times$10$^{53}$~erg \citep{Wang2019,Li2023a}. The prompt emission light curve consists of three well-separated emission pulses. The first emission episode (i.e., $P_1$) starts at $T_0$ and lasts for $\sim$2.35~s, the second emission episode (i.e., $P_2$) exhibits multiple peaks and lasts from $T_0$+2.35~s to $T_0$+15~s and slightly overlapping with $P_{1}$, and the significantly fainter third emission episode (i.e., $P_3$) extends from $T_0$+15~s to $T_0$+25~s. The first two emission episodes have very hard the low-energy photon $\alpha$ indices, with the majority of the $\alpha$ indices in $P_1$ and $P_2$ beyond the line-of-death of synchrotron emission \citep[-2/3,][]{Preece1998}, while the third emission episode pulse has relatively soft $\alpha$ indices, indicating a transition from fireball (thermal, $P_1$ and $P_2$) to Poynting-flux-dominated (nonthermal, $P_3$) outflow. The afterglow emission measured by {\it Swift}-XRT starts at $\sim$ $T_{0}$+68~s. The first GeV photon was observed by {\it Fermi}-LAT at $T_{0}$+2.1~s, and the highest-energy photon is a 22.9~GeV event observed at $T_{0}$+15~s \citep{Wang2019}.

Thermal photons of the prompt emission generated by the photosphere are recognized as one of the leading radiative process in gamma-ray burst (GRB) physics \cite[e.g.,][]{Goodman1986,Paczynski1986,Peer2007}. The violation of the synchrotron emission limit encourages us to search for an additional thermal component \citep{Li2023a}. A highly detailed time-integrated and time-resolved spectral analysis of GRB 190114C has been presented in a recent study \citep{Li2023a}, and its high fluence ($\sim$ 4.436$\times$10$^{-4}$ erg cm$^{-2}$) allows us to conduct the time-resolved spectral analysis in great detail and to study their variations down to a very short time-scale ($\sim$0.1 s) while preserving a high significance. By carrying out the detailed spectral analysis for GRB 190114C, \cite{Li2023a} reported that the spectra in both $P_1$ and $P_2$ exhibit significant deviation from a single Band function \citep{Band1993} and can be best fitted using two components: a subdominant thermal blackbody (BB) component \citep[e.g.,][]{Ryde2004,Guiriec2015a,Li2023c}, accompanied by a dominant nonthermal (Band-like) component, suggesting the existence of a strong thermal component. Indeed, the thermal component confidently presents in the time bins from $T_{0}$+0.55~s to $T_{0}$+1.93~s in $P_1$ and from $T_{0}$+2.45~s to $T_{0}$+5.69~s in $P_2$, covering the peaks of the initial two emission pulses, and precisely corresponds to the episodes where the $\alpha$ indices are beyond the synchrotron limit \citep{Li2023a}. More interestingly, the two thermal components between the well-separated thermal pulses ($P_1$ and $P_2$) evolve independently, as inferred from their observational and physical parameters derived from the standard fireball model \citep{Peer2007}. A large percentage of $\sim$20\% of the energy of thermal emission is present in the $\gamma$-ray prompt emission, making it one of the most thermal-prominent \emph{Fermi} GRBs. The strong and independent pulse-wise thermal components observed in the time-resolved spectral analysis of GRB 190114C make it a good case to study the photosphere properties, allowing us for the first time to study a fine time-resolved spectral analysis and track the BB evolution among the different pulses in a single GRB. These observational features found in GRB 190114C strongly support the evidence of a shell-like structure during the prompt emission phase and provide a good opportunity to study GRB ejecta composition and the efficiency of GRB radiation as well.

The radiative efficiency of a burst is another interesting subject related to the GRB prompt emission mechanism, which describes how efficiently the jet converts its energy to radiation. GRB radiative efficiency can be defined as \citep{Lloyd-Ronning2004}
\begin{eqnarray}
    \eta_\gamma & \equiv & \frac{E_\gamma}{E_{\rm tot}} = \frac{E_\gamma}{E_\gamma+E_k} = \frac{L_\gamma}{L_{w,0}}, \label{eq:eta_gam} \end{eqnarray}
where $E_\gamma$, $E_k$, and $E_{\rm tot}$ are isotropic-equivalent $\gamma$-ray energy, afterglow kinetic energy, and total energy, respectively, and $L_\gamma$ and $L_{w,0}$ are the isotropic-equivalent average $\gamma$-ray luminosity and total wind luminosity at the central engine, respectively. In order to calculate the radiative efficiency ($\eta_\gamma$) of a GRB, according to Equation (\ref{eq:eta_gam}), one needs to know the isotropic-equivalent $\gamma$-ray energy $E_\gamma$ and the blastwave kinetic energy $E_{\rm K}$. The $E_\gamma$ can be directly measured using spectral parameters. The $E_{\rm k}$ term, on the other hand, cannot be directly measured from observations. The traditional method \citep[e.g.,][]{Zhang2007b,Wang2015} used the afterglow data through modeling to estimate its value, but the estimated value typically carries large uncertainties since it depends on many uncertain shock microphysics parameters, primarily $\epsilon_e$ \citep{freedman01}, but also $\epsilon_{\rm B}$ and electron spectral index $p$ \citep{Zhang2007b,Wang2015}, where $\epsilon_{e}$ and $\epsilon_{\rm B}$ are the fraction of the shocked energy density transferred to the magnetic fields and electrons, respectively. This resulted in high uncertainties in the derived GRB radiative efficiency, ranging from below 10\% to more than 90\% \citep{Zhang2007b,Wang2015,Li2018b}. By combining the prompt emission photosphere emission data and early afterglow data, \cite{Zhang2021} proposed a new method to directly dissect the GRB fireball energy budget into three components and measure their values. As a result, GRB radiation efficiency can also be directly calculated with little uncertainty. The method requires a GRB observable with a dominant thermal spectral component, a deceleration bump feature in the early afterglow light curve, and a measured redshift. The measured parameters include the initial dimensionless specific enthalpy ($\eta$), bulk Lorentz factors at the photosphere radius ($\Gamma_{\rm ph}$), and before fireball deceleration ($\Gamma_{0}$), the amount of mass loading ($M$), and GRB radiative efficiency ($\eta_\gamma$). These measured parameters only weakly depend on the density $n$ of the interstellar medium when the composition ${\cal Y}$ parameter (typically unity) is specified. Once these fireball parameters can be precisely measured, one can also estimate the blastwave kinetic energy as $E_{\rm K}=\Gamma_0 M c^{2}$. As a result, GRB radiative efficiency $\eta_\gamma\equiv E_{\gamma}/(E_{\gamma}+E_{\rm K})$ can be also derived.

By combining the prompt emission and assuming that the afterglow emission starts at $\sim$ 6~s as supported by several independent studies in the literature \citep[e.g.,][]{MAGICCollaboration2019,2019A&A...626A..12R,ajello2020, Ursi2020}, \cite{Li2023a} first applied this new method \citep{Zhang2021} to GRB 190114C, and directly measured all the parameters for the first time (see Table 6 in \citealt{Li2023a}). The measured parameters at different emission sites are consistent with the expectation of a fireball, and such a feature clearly exhibits a whole picture of self-consistency from early prompt to late afterglow emission. GRB 190114C, therefore, exhibits the evolution of a textbook relativistic fireball. In this paper, we show that $P_3$ could also be explained as originating from the synchrotron self-Compton (SSC) pulse from a reverse shock (RS) afterglow emission, indicating the onset signature of afterglow emission \citep{Fraija2019}. Using the strong thermal components observed in $P_1$ and $P_2$, along with the early afterglow data from $P_3$, we can recalculate all parameters. On the other hand, we may also perform a similar pulse-wise analysis if we take into account the distinct thermal components that were seen in $P_1$ and $P_2$ and make the assumption that the emission before and after $P_3$ may represent the afterglow emission component corresponding to the two initial pulses $P_1$ and $P_2$.

The paper is organized as follows. The spectral analysis is presented in Section \ref{sec:SpecFit}. The methodology is applied to GRB 190114C based on its global and pulse-wise properties and is presented in Section \ref{sec:Results}. A discussion is presented in Section \ref{sec:Discussion} and our conclusions are summarized in Section \ref{sec:Conclusion}. Throughout the paper, the standard $\Lambda$-CDM cosmology with the parameters $H_{0}= 67.4$ ${\rm km s^{-1}}$ ${\rm Mpc^{-1}}$, $\Omega_{M}=0.315$, and $\Omega_{\Lambda}=0.685$ are adopted \citep{PlanckCollaboration2018}. 

\section{Time-integrated and Time-resolved Spectral Analysis}\label{sec:SpecFit}

We first perform the time-integrated spectral analysis with the entire duration ($T_{90}$) of the burst (i.e., from $T_0$ + 0~s to $T_0$ + 116~s). The background is fitted with polynomial functions with the determined polynomial order (0-4) by applying a likelihood ratio test, using the two off-source time intervals, the intervals pre- (-20~s to -10~s) and post- (180~s to 200~s) the burst. We select two brightest NaI detectors (n3, n4) to obtain an angle of incidence less than 60 degrees \citep{Goldstein2012}, as well as one BGO detector (b0) with the lowest angle of incidence. The Time-Tagged Event (TTE) data type is used for NaI data (8 keV - 1 MeV) and BGO data (200 keV - 40 MeV). Following the standard practice \citep{Li2019b,Li2019c,Li2019a,Yu2019,Burgess2019,Li2020,Li2021a,Li2021b,Li2023c,Li2022} provided by the \emph{Fermi} Science Term, the spectral analysis is carried out by using the Bayesian iterations of Markov Chain Monte Carlo, which are performed by a Python package {\it Multi-Mission Maximum Likelihood Framework} (3ML, \citealt{Vianello2015}). To search for the best model representing the spectral shape, we first fit the spectrum by examining various frequently used spectral models by performing a detailed spectral analysis and model comparisons. The models utilized include a power law (PL), cutoff power law (CPL), Band, smoothly broken power law (SBKL), PL+BB, CPL+BB, PL+bandcut, and Band+BB, respectively\footnote{We use the following abbreviations: simple power-law (PL), CPL (cutoff power-law), smoothly broken power law (SBKPL), and BB (blackbody).}. Our refined spectral analysis suggests that the CPL+BB (or the Band+BB) fitting is much better than the others, and its AIC and BIC scores are at least dozens of points lower than other models. This is quite surprising, and indicates that adding a thermal component largely improves the spectral fitting.  

\begin{figure*}
\includegraphics[width=1.0\hsize,clip]{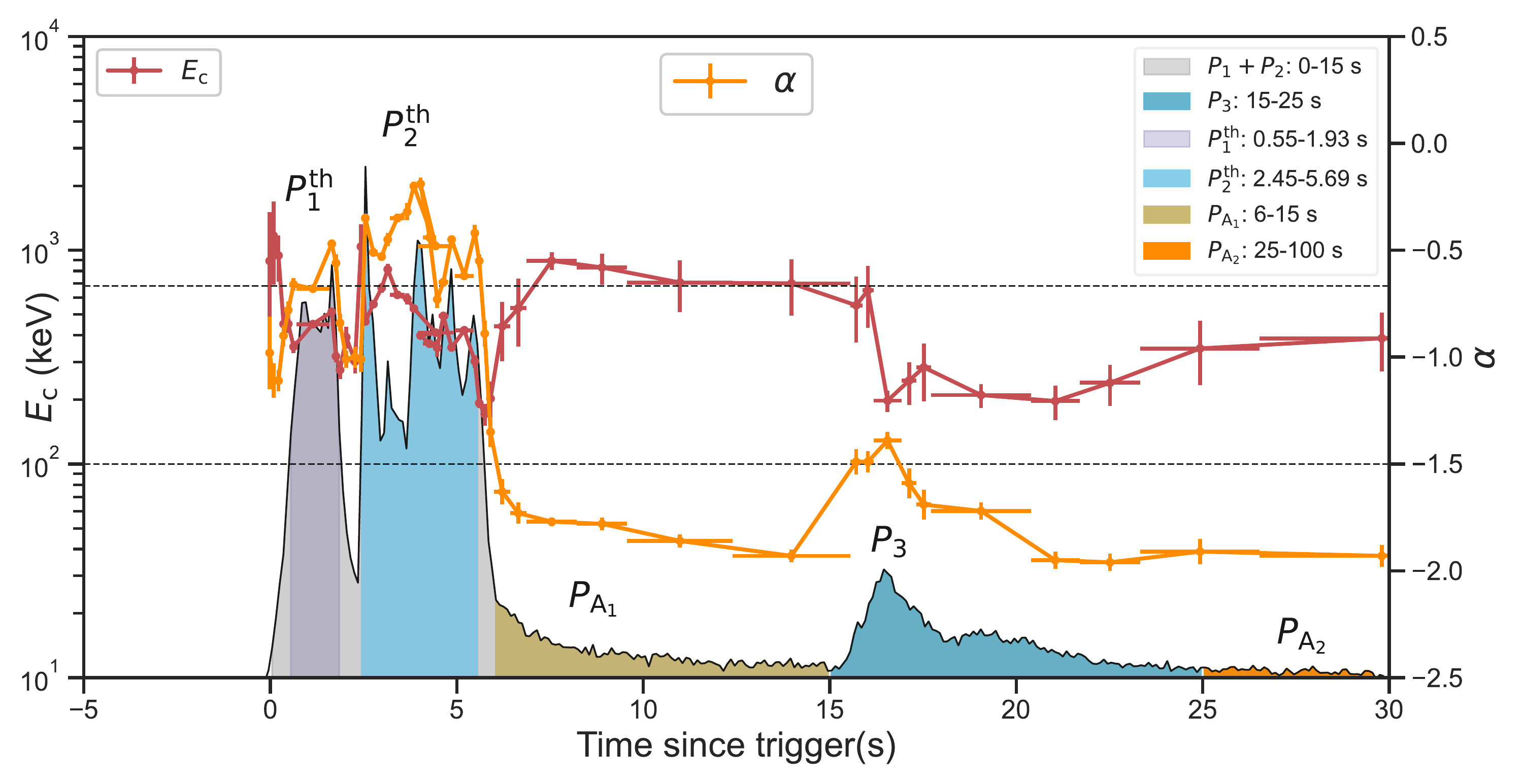}   
\caption{The GBM count light curve during the time span of $0-30$~s. The data points connected by solid lines in orange and red represent the temporal evolution of cutoff energy $E_{\rm c}$ (left panel) and $\alpha$ (right panel) from the CPL-alone model fits. Two horizontal dashed lines represent the limiting values of $\alpha$=-2/3 and $\alpha$=-3/2 for electrons in the synchrotron slow- and fast-cooling regimes, respectively. The shaded regions marked with different colors denote the different episodes: the entire duration of $P_{1}$+$P_{2}$ (gray) and the entire duration of $P_{3}$ (cyan) for the case based on global properties (see Section \ref{subsec:Global}), and two independent thermal emission $P^{\rm th}_{1}$ (magnetic) and $P^{\rm th}_{2}$ (sky blue) and corresponding two afterglow emission $P_{\rm A_{1}}$ (yellow) and $P_{\rm A_{2}}$ (dark orange) for the case based on pulse-wise properties (see Section \ref{subsec:PulseWise}).}
\label{fig:Global}
\end{figure*}

We further conduct the time-resolved spectral analysis from -1.0~s to 150~s covering the $T_{90}$ duration. The time bins are selected by using the Bayesian blocks method (BBlocks, \citealt{Scargle2013}) to TTE light curve of the most strongly illuminated GBM detector (n4), while other detectors used are binned in matching time bins. In total, 49 time bins of spectra are obtained. We first use the typical GRB spectral model the CPL model to fit the data, and to obtain the temporal evolution of the low energy photon index $\alpha$. The first result we find is that the temporal evolution of the PL photon index $\alpha$ (below the spectral energies peak) obtained from the CPL model is very unusual. It is very hard in the first 6~s, covering the peaks of the first two pulses and significantly violating the optically thin synchrotron limit \citep{Preece1998}, then it crosses the synchrotron limit with a hard-to-soft trend, and several spectra have $\alpha$ close to 0 (see Figure \ref{fig:Global}). Eventually, it is softer than the fast-cooling limit of -3/2 \citep{Preece1998}. The break of the fast-cooling limit at a later time infers the GRB has entered the afterglow phase, and the cutoff energy is found to be $<$100 keV after 50~s. Observing such a wide span of photon index in a single GRB is unique, especially since the breaks occur in sequences of time bins and are confirmed by data of high statistical significance ($S>$20). The second result we find is that a time-integrated fitting of the whole $T_{90}$ shows the consistency; $\sim$20\% of the {\it Fermi}-GBM $\gamma$-ray energy is in the thermal emission, and the average temperature is greater than 130 keV. This high percentage of thermal flux reminds in the GRB 090902B, which is famous for the intense thermal emission appearing in the first half of the prompt emission. Such a value is believed to be one of the highest on record for thermal emission in the hybrid spectral case (a dominat nonthermal emission component with a subdominat thermal component). 

\section{Directly Deriving the fireball parameters and radiative efficiency}\label{sec:Results}

\subsection{The Method} 

\cite{Zhang2021} proposed a new method to directly dissect the GRB fireball energy budget into three components and measure their values. The method requires a GRB observable with a dominant thermal spectral component, a deceleration bump feature in the early afterglow light curve, and a measured redshift. The measured parameters include the initial dimensionless specific enthalpy ($\eta$), bulk Lorentz factors at the photosphere radius ($\Gamma_{\rm ph}$), and before fireball deceleration ($\Gamma_{0}$), the amount of mass loading ($M$), and GRB radiative efficiency ($\eta_\gamma$). These measured parameters only weakly depend on the density $n$ of the interstellar medium when the composition ${\cal Y}$ parameter (typically unity) is specified. Following is a brief description of the relevant calculations.

The initial, total energy of a fireball is
\begin{equation}
E_{\rm tot} = \eta M c^2.
\label{eq:Etot}
\end{equation}
The fireball undergoes rapid acceleration and reaches a Lorentz factor $\Gamma_{\rm ph}$ at the photosphere. 
The internal energy released as thermal emission can be estimated as
\begin{equation}
E_{\rm th} = (\eta-\Gamma_{\rm ph}) M c^2,
\label{eq:Eth}
\end{equation}
Afterwards, the fireball moves at an almost constant speed until internal dissipation at internal shocks occurs at a larger distance.
The emitted nonthermal emission can be estimated as
\begin{equation}
E_{\rm nth} = (\Gamma_{\rm ph}-\Gamma_0) M c^2,
\label{eq:Enth}
\end{equation}
where $\Gamma_0$ is the Lorentz factor after the dissipation and the initial Lorentz factor in the afterglow phase. 

The Lorentz factor at photosphere radius $\Gamma_{\rm ph}$ can be estimated as (modified from \cite{Peer2007, Begue2014}, see \cite{Zhang2021} for details)
\begin{equation}
\begin{split}
    \Gamma_{\rm ph} 
     &=  \left[(1+z)^2 D_{\rm L} \frac{ {\cal Y}\sigma_{\rm T} F^{\rm obs}_\gamma}{2 m_p c^3 {\cal R}} \frac{\eta^{3/2}}{\eta-\Gamma_0} \right]^{2/9}, \\
{\cal R}&=\left(\frac{F^{\rm obs}_{\rm BB}}{\sigma_{\rm B} T^{4}}\right)^{1/2}.
\end{split}
\label{eq:Gammaph_new}
\end{equation}
which involves several direct observables including redshift $z$, total flux $F^{\rm obs}_\gamma$, thermal flux $F^{\rm obs}_{\rm BB}$, and the observed temperature $T$. Other parameters include the pair multiplicity parameter ${\cal Y}$, which is commonly taken as $1$, the luminosity distance $D_{\rm L}$ computed from the redshift adopting the the Friedmann-Lemaitre-Robertson-Walker (FLRW) cosmology, and fundamental constants such as the speed of light $c$, proton mass $m_{\rm p}$, Thomson cross section $\sigma_{\rm T}$, and Stefan-Boltzmann constant $\sigma_{\rm B}$.

The initial Lorentz factor of the afterglow phase $\Gamma_0$ can be derived by equating the kinetic energy to the swept-up ISM mass at the deceleration time $t_{\rm dec}$, which is an observable indicated by a light-curve pulse (the third pulse for 190114C). Using Equation (7.81) of \cite{Zhang2018} and the above arguments, we derive
\begin{equation}
\begin{split}
\Gamma_{0}\simeq0.9^{3/8} \left( \frac{3E_{\rm k}(1+z)^{3}}{2\Pi \hat{\gamma} n m_{\rm p} c^{5} t^{3}_{\rm dec}} \right)^{1/8}\\
\simeq 170 t_{\rm dec,2}^{-3/8} \left(\frac{1+z}{2}\right)^{3/8} \left(\frac{E_{\gamma,52}}{n}\right)^{1/8} \left(\frac{\Gamma_0}{\eta-\Gamma_0} \right)^{1/8}.
\label{eq:Gamma0}
\end{split}
\end{equation}
where $\hat{\gamma}$(=4/3) is the numerical coefficient, $n$ is the ISM density assumed as one particle per cubic centimetre as usual, $t_{\rm dec}$ is the peak time of the third pulse since we assume that this point has already in the deceleration time, $E_{\rm k}$ is the isotropic blastwave kinetic energy $E_{\rm K,iso}$ as we have calculated above, and $E_{\gamma,52}=E_{\rm th,52}+E_{\rm nth,52}$.

Simultaneously solving Equations.(\ref{eq:Eth}--\ref{eq:Gamma0}), we obtain the fireball parameters $\eta$, $\Gamma_{\rm ph}$, $M$ and $\Gamma_{0}$, and in turn, we can calculate the kinetic energy of the afterglow 
\begin{equation}
E_k = \Gamma_0 M c^2,
\end{equation}
and the efficiency of the prompt $\gamma$-ray emission (see Table \ref{tab:global}) 
\begin{equation}
\eta_\gamma = \frac{E_{\rm th}+E_{\rm nth}}{E_{\rm tot}} = \frac{\eta-\Gamma_0}{\eta}. 
\label{eq:eta_gam2}
\end{equation}

\subsection{Global properties}\label{subsec:Global}

\subsubsection{The Deceleration of Pulse $P_{3}$: SSC Pulse from the RS Afterglow Emission} \label{subsec:P2} 

The shell-merged fireball continues to move approximately at the Lorentz factor $\Gamma_0$ and then decelerates when the swept-up ISM mass equals to its kinetic energy. This moment denotes the starting of the afterglow phase, accompanied by a bulging pulse on the light curve and the softening of the spectral index. This apparently coincides with the third weak pulse $P_{3}$ of which the time-resolved spectral indices (see Figure \ref{fig:Global}) ($\alpha \sim$ -1.5 to -2) are much softer than those in the first two pulses ($P_{1}$ and $P_{2}$), but are consistent with the typical values of synchrotron radiation of afterglow origin. 

Considering that the observational properties in $P_{3}$ are likely to be consistent with the theoretical model of SSC emission from the RS \citep[e.g.,][]{Meszaros1997,Sari1999b,Kobayashi2003,Wei2003}, we try to check the interpretation that the emission of the third $P_{3}$ pulse (during the time intervals between 15~s and 25~s) as from the optical photons of the RS upscattered to the X-rays and $\gamma$-rays. 

Our approach assumes that the shell is in a certain regime, and then applies our method to calculate the Lorentz factors, kinetic energy, and other parameters as listed in Table \ref{tab:global}. We then use the calculated parameters to obtain the characteristic values for determining the regime of the shell, checking for consistency. Our results show that choosing the thick-shell regime makes the entire process consistent.

First, we judge from the critical Lorentz factor:
\begin{equation}
\Gamma_c\equiv  \left(\frac{3E_{\rm K}\,(1+z)^3}{32\pi m_p c^5\, n}\right)^{\frac{1}{8}}  T_{\rm prompt}^{-\frac38} = 386.
\end{equation}
Here we adopt the parameter of $T_{\rm prompt}=15$~s as the ending time of the second pulse. A longer $T_{\rm prompt}$ brings a smaller $\Gamma_{\rm c}$. If we choose the $T_{\rm prompt} = T_{90} = 116$~s from the Fermi-GBM, the resulting critical Lorentz factor will be reduced to 179. Therefore, $\Gamma_{\rm c}$ is smaller than the Lorentz factor after the deceleration time $\Gamma_0 = 507$ and much smaller than the initial Lorentz factor of $\eta = 708$, indicating the thick-regime case.

Second, the Sedov length ($l$), the shock crossing time ($t_{\rm x}$) and the observed shell width ($\Delta$) can be obtained by the computed kinetic energy $E_{\rm k}$:
\begin{equation}
l = (\frac{3E_{\rm k}}{4\pi n m_{p} c^2})^\frac13 = 4.98 \times 10^{18}~{\rm cm}
\end{equation}
\begin{equation}
t_{\rm x} = \Gamma_0^{-\frac83} \left(\frac{3 E_{\rm k} (1+z)^3 }{32\pi m_p c^5 n}\right)^\frac13 = 7.24~{\rm s},
\end{equation}
\begin{equation}
\Delta=2c (1+z)^{-1}\,t_{\rm x} = 3.04 \times 10^{11}~cm.
\end{equation}
The obtained $t_{\rm x}$ is smaller than the $T_{90}$, confirming the thick-shell regime.

We also notice that shell width $\Delta$ is only about two times of $ l/2\Gamma_0^{8/3} = 1.52 \times 10^{11}$~cm, and this is in line with $P_{3}$ being a faint pulse. Its energy released is $E_{\gamma, \rm iso}\approx 1.20 \times 10^{52}$ erg, taking up a very small portion ($1.5\%$) of the kinetic energy $E_{\rm k,iso}$ = (7.8$\pm$0.6) $\times$ 10$^{53}$ erg, which enables us to simplify the deceleration Lorentz factor $\Gamma_{\rm d} \simeq \Gamma_0$.

The ratio of electron and magnetic field equipartition parameters $\epsilon_{\rm e}/\epsilon_{\rm B} = 132 \gg 1$ indicates an SSC-dominated regime, where $\epsilon_{\rm B}$ and $\epsilon_{\rm e}$ are the fractions of the shock energy transferred to the magnetic field and electrons, respectively. For an RS, with $\Gamma_{\rm d} > 500$, the first-order SSC normally upscatters the photons to the keV/MeV, and the second-order IC boosts photons again to the GeV range. Observationally, the luminosity of photons above the GeV range is below the keV/MeV luminosity, suggesting that the second-order SSC is suppressed by the Klein-Nishina effect. Hence, we may consider only the first-order IC and derive the ratio of Compton scattering parameter \citep{Sari2001} $x = (\epsilon_{\rm e}/\epsilon_{\rm B})^{1/2} =12$.

RS has been widely studies, for similar previous cases and the related theories, we refer to \citet{2001ApJ...546L..33W,2001ApJ...556.1010W,2003A&A...402L...9W,2007ApJ...655..391K,2014ApJ...789..146U,Fraija2020} and the reference therein. Our applied model generally follows the model in \citet{2005A&A...439..957W}, in which the RS SSC emission is modeled and applied, and the equations are mainly adopted from \citet{2003A&A...402L...9W,2005A&A...439..957W}.

A contact discontinuity separates the reverse and forward shocks, maintaining equal pressure on both sides. Hence, the typical frequencies and the flux from the synchrotron emission of the RS are connected to the corresponding values of the forward shock. Assuming the magnetic fields are the same in both shocked, we have
\begin{equation}
\begin{split}
    \nu_{\rm m, r} &\simeq \Gamma_{\rm d}^{-2} \nu_{\rm m, f} \\
&= 3.89\times 10^{15}\left(1+z\right)^{1/2} (\frac{p-2}{p-1})^2 \Gamma_{\rm d,3}^{2} \epsilon_{\rm B, -2}^{1/2}\epsilon_{\rm e, -1}^2 n_0^{1/2} \;\;\; {\rm Hz} \\
&= 0.62  \;\;\; {\rm eV} 
\end{split}
\end{equation}

\begin{equation}
\begin{split}
    \nu_{\rm c, r} &= \nu_{\rm c, f} \\ &= 7.94\times
10^{16}(1+Y^{\rm IC})^{-2}\left(1+z\right)^{-1/2}\epsilon_{\rm
B,-2}^{-3/2}E_{k, 52}^{-1/2}n_0^{-1}t_{\rm dec,2}^{-1/2} \;\;\; {\rm Hz} \\
&= 0.96  \;\;\; {\rm keV}
\end{split}
\end{equation}
where $f$ and $r$ denote forward and reverse respectively, and $Y^{\rm IC}$ is the inverse Compton (IC) parameter and the IC parameter $Y^{\rm IC}(=E_{\rm GeV}/E_{\rm MeV}$) constrained from the observations in GRB 190114C, e.g. $Y^{\rm IC}=0.75$~\citep[e.g.,][]{Wang2019b}.
The cooling frequency is much larger than the minimal frequency ($\nu_{\rm c, r} \gg \nu_{\rm m, r}$). These frequencies are upscattered to
\begin{eqnarray}
\nu_{\rm m,r}^{\rm IC} &\simeq& \gamma_m^2 \nu_{m,r}  = 13 ~\mbox{keV}, \\
\nu_{\rm c,r}^{\rm IC} &\simeq& \gamma_c^2 \nu_{c,r}  = 91 ~\mbox{MeV},
\end{eqnarray}
where
\begin{eqnarray}
\gamma_m &\simeq& \left(\frac{p-2}{p-1} \right) \epsilon_{e} \left(\frac{m_p}{m_e}\right) = 145, \\
\gamma_c &\simeq& \gamma_m \left(\frac{\nu_c}{\nu_m}\right)^{1/2} = 9843
\end{eqnarray}
Here $m_{\rm p}$ and $m_{\rm e}$ are the masses of the proton and the electron, respectively. The flux between $\nu_{\rm m}^{\rm IC}$ and $\nu_{\rm c}^{\rm IC}$ drops at $\sim t^{-(3p+1)/3}$, where $p$ is the PL index of the electron distribution, from the \textit{Fermi}-GBM observations, the flux drops at $\sim t^{-3.3}$, so $p=2.9$ is derived and applied to obtain the above $\nu_{\rm m}$ and $\nu_{\rm c}$ values. The flux density at the $\nu_{\rm m}$ is given by
\begin{equation}
\begin{split}
    f_{\rm \nu_m, r} &\simeq \Gamma_d f_{\rm \nu_m, f} \\ &= 11 \left(1+z\right) \Gamma_{d,3} \epsilon_{\rm B, -2}^{1/2}E_{k, 52} n_0^{-1} D_{\rm L, 28}^{-2}\;\;\; {\rm Jy} \\
&= 185  \;\;\; {\rm Jy} 
\end{split}
\end{equation}
where $D_{\rm L}$ is the luminosity distance of the source. This corresponds to the SSC flux 
\begin{equation}
\begin{split}
    f_{\rm r}^{\rm IC} = \nu_{\rm m,r}^{\rm IC} f_{\nu_{\rm m,r}^{\rm IC}} \simeq x \nu_{\rm m,r} f_{\nu_{\rm m,r}} = 5.7\times10^{-6}   \;\;\; {\rm erg~cm^{-2}~s^{-1} }
\end{split}
\label{eq:reverse-flux}
\end{equation}

The above results suggest that the modeled SSC emission from the RS in the keV range and the flux in the order of $10^{-6}$~erg~cm$^{-2}$~s$^{-1}$ is consistent with observation of $P_3$.

The optical afterglow data also offer support to this interpretation. Even though there is no early optical data before $\sim 33$~s, the optical light-curve fitting requires a steeper slope ($\sim -1.76$) early on before transitioning to a normal slope ($\sim -1$), consistent with the superposition of an RS component with the forward shock component\footnote{Throughout the paper, the convention $F_{t,\nu}=t^{-\alpha}\nu^{-\beta}$  is applied, where $\alpha$ and $\beta$ are the temporal and spectral indices, respectively.}.

\begin{figure*}
\centering
\includegraphics[width=1.0\hsize,clip]{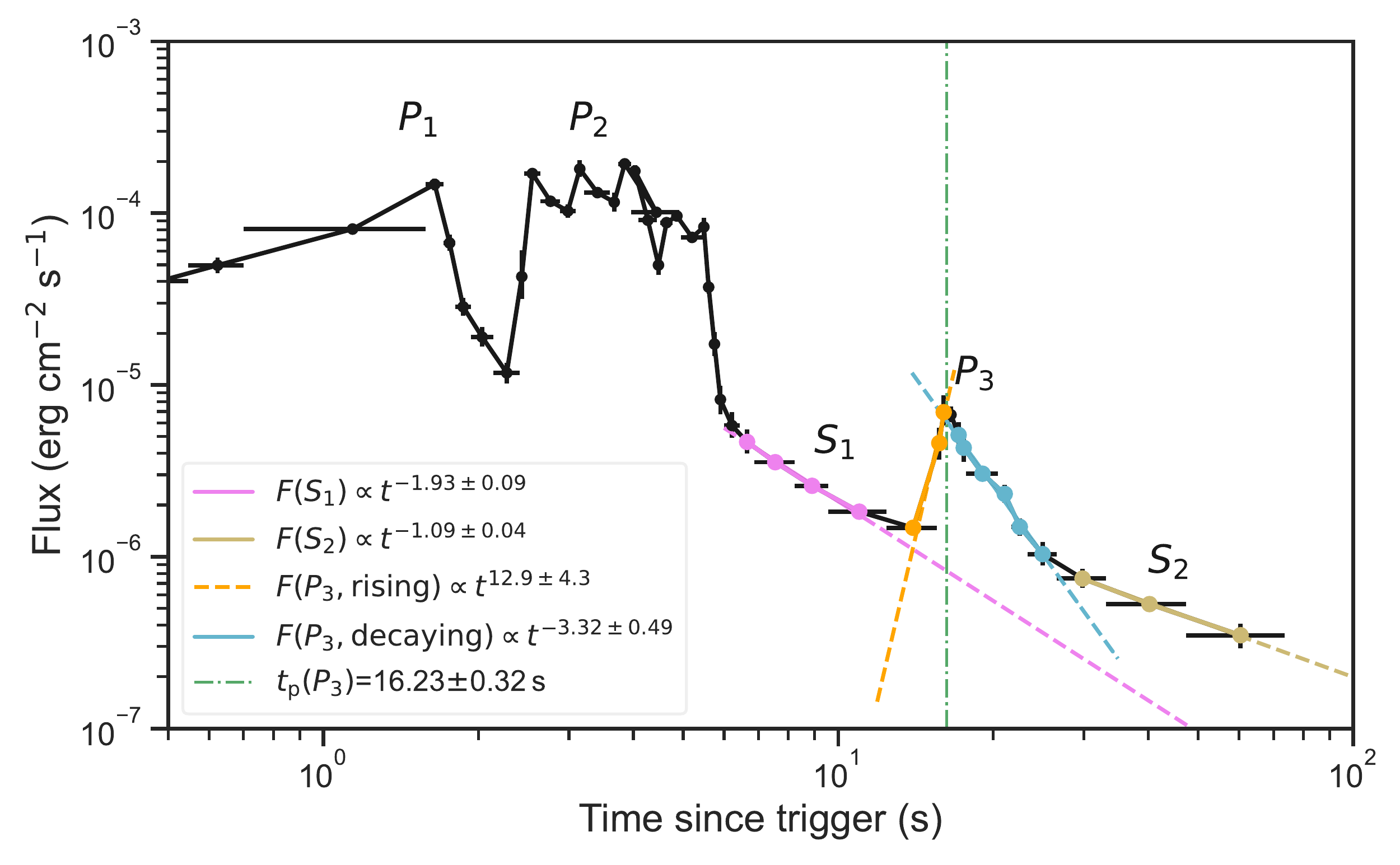} 
\caption{The GBM energy light curve (black) with the best fitting to $S_{1}$ (violet), $S_{2}$ (yellow), and the rising (orange) and decay (cyan) phases of $P_{3}$ using the PL model. The value of $t_{\rm p}(P_{3})$ (the green vertical dashed line) is used to estimate $\Gamma$ in Equation (\ref{eq:Gamma0}) while the value of $\hat{\alpha}(P_{3})=-3.32\pm0.49$ denotes the temporal decay index of the RS from the afterglow emission. The decay index of the afterglow emission from $P_{1}$ is $\hat{\alpha}(S_{1})=-1.93\pm0.09$, which is significantly steeper than a typical value for afterglow emission measured from other GRBs. This is because part of the energy flux in this segment has clearly been contributed from $P_{2}$, whereas that from $P_{2}$ is $\hat{\alpha}(S_{2})=-1.09\pm0.04$, which is in good agreement with typical values observed from afterglow emission.}\label{fig:p3_peak}
\end{figure*}

In Figure \ref{fig:p3_peak}, we show the log ($F$)-log($t-t_{0}$) plots with a temporal fit to the decay phase of the third peak using the single PL model (typical model for afterglow emission), setting $t_{0}$ at the trigger time log($t-t_{\rm trigger}$), which is physically motivated to study external-shock-powered light curves \citep{2007ApJ...655..973K}. We get $F\propto t^{-3.32\pm0.49}$~s. The obtained decay index ($-3.32\pm0.49$) in this way is much steeper than the typical values observed from a forward shock emission in afterglow, but is in good agreement with an RS afterglow emission predicted by the model discussed above. Therefore, $P_3$ could be originated from an RS.

In addition, the above computation takes $n=1$, and the adopted parameters such as $\epsilon_{\rm e}$, $\epsilon_{\rm B}$ and $E_{\rm k}$ are all derived from $n=1$. Hence, the resulting consistency with the observation of $P_3$ proves that $n$ is indeed $\sim$ 1.

To clarify, the equipartition parameters are not necessarily the same in the reverse and forward shocks. These parameters are typically assumed or derived from fitting the afterglow data. In our study, the equipartition parameters for the forward shock are derived from observations, but there is no effective method yet for obtaining them from observations for the RS.  Prior research, such as \citet{2012ApJ...755...12V}, has posited that the equipartition parameters in the reverse and forward shocks are identical, while others, like \citet{2012ApJ...751...33F}, have suggested they are independent. In this article, since we have derived values from the forward shock, the most natural assumption in the absence of definitive proof of their consistency is to consider them as identical. In fact, if we assume they are not the same, based on past findings \citep{2005ApJ...628..315Z, 2008ApJ...687..443G}, the $\epsilon_{\rm B}$ of the RS is higher than that of the forward shock. If we assume it is 10 times larger ($\epsilon_{\rm B} \sim 0.01$) in the RS than in the forward shock, according to Equation \ref{eq:reverse-flux}, $f_{\rm r}^{\rm IC} \propto \epsilon_B^{1/2}$,  the radiative flux of the RS would be three times higher. However, this assumption does not significantly alter our estimations and conclusions.

\subsubsection{Results} 

Following the physical picture that the central engine releases two sequences of energy forming $P_{1}$ and $P_{2}$, which merge and enter the afterglow phase generating $P_3$, we can directly derive the fireball parameters and GRB radiative efficiency for GRB 190114C following the method described in \cite{Zhang2021}. We first perform a spectral fit using the CPL+BB model (see Section \ref{sec:SpecFit}) by treating both thermal components observed in $P_{1}$ and $P_{2}$ as a whole. The observed parameters obtained from the spectral analysis, therefore, include the isotropic-equivalent thermal energy $E_{\rm th,iso}$ and isotropic-equivalent nonthermal energy $E_{\rm nth,iso}$, the thermal $F^{\rm obs}_{\rm BB}$ and total $F^{\rm obs}_{\gamma}$ energy flux, and the average temperature $kT^{\rm obs}$ (see Table \ref{tab:global}). The deceleration time $t_{\rm dec}$ ($t_{\rm p}$) is measured from the third pulse assuming it originates from the SSC pulse from the RS afterglow emission (see Section \ref{subsec:P2}). The peak time ($t_{\rm p}$) is determined using the fast-rising and exponential decay model \citep{Kocevski2003} to fit the 1024 ms counts light curve of the third pulse (see Table \ref{tab:global}). By combining the thermal component in the first two pulses with a non-thermal component in the third pulse (see Figure \ref{fig:Global}) and following the method proposed in \cite{Zhang2021}, one can directly determine the fireball characteristics. The measured parameters include the initial dimensionless specific enthalpy ($\eta$), bulk Lorentz factors at the photosphere radius ($\Gamma_{\rm ph}$), and before fireball deceleration ($\Gamma_{0}$), the amount of mass loading ($M$), the kinetic energy in the fireball $E_{\rm K}$, and GRB radiative efficiency ($\eta_{\gamma}$).

In this scenario, the prompt emission of GRB 190114C lasts from $0$~s to $15$~s, and thermal emission is prominent during $0.55$~s-$1.93$~s and $2.45$~s-$5.69$~s. With the observed properties (see the upper panel of Table \ref{tab:global}), and substituting all these values in Equations (\ref{eq:Etot}-\ref{eq:eta_gam2}), we obtain the fireball characteristics (see the middle panel of Table \ref{tab:global}), where $M_\odot$ is the mass of the sun ($1.9891 \times10^{33} \rm ~g$). The derived fireball parameters consist of the dimensionless specific enthalpy at the engine $\eta$, the bulk Lorentz factor at the site of the photosphere $\Gamma_{\rm ph}$, the initial afterglow Lorentz factor before the deceleration phase $\Gamma_{0}$, the isotropic-equivalent total mass $M_{\rm iso}$, the kinetic energy in the fireball $E_{\rm k,iso}$, and the $\gamma$-ray radiative efficiency $\eta_\gamma$, as well as the fraction of the shocked energy density transferred to the magnetic fields ($\epsilon_{\rm B}$) and electrons ($\epsilon_{e}$), the characteristic synchrotron frequency ($\nu_{\rm m}$) and the cooling frequency ($\nu_{\rm c}$) of minimum-energy injected electrons, and the Klein-Nishina frequency ($\nu_{\rm KN}$). The measured quantities from observations and the derived fireball parameters using our new methods with assuming ${\cal Y}=1$ and $n=1~cm^{-3}$. We note that (1) we assume that at the time of $P_{3}$ the swept-up mass from the environment is much less than the mass ejected from the central engine, this assumption is validated by the following numerical result. The mass ejected from the black hole ($M$= 8.6$\times$10$^{-4}$$M_{\sun}$) is much more massive than the mass swept up from the ISM within the first 20~s ($\sim 10^{-4}M_{\sun}$), which verifies the feasibility of ignoring the swept-up mass during the above modeling. (2) A relevantly high GRB radiative efficiency is obtained, $\eta_\gamma=(28.3\pm1.4)\%$, suggesting that if GRBs are powered by fireballs, the efficiency can sometimes be high.

\subsection{Pulse-wise Properties}\label{subsec:PulseWise}  

\subsubsection{Independent Thermal Pulses: The First Pulse $P_1$ and the Second Pulse $P_2$}

The traditional method to derive the photosphere properties invokes the standard fireball model \citep{Meszaros2000, Peer2015,Zhang2018}. Within this framework, the fireball invokes thermally accelerated, matter-dominated, and finally shocks-decelerated ejecta \citep{Goodman1986,Paczynski1986}. The identification of the strong thermal component in GRB 190114C allows us to determine the physical properties of the relativistic outflow within the framework of the non-dissipative photosphere theory \citep{Peer2007,vereshchagin2017}. 

The observational and derived photosphere parameters (see Section \ref{subsec:A1} for details), including the BB temperature ($T$), the effective radius parameter ($\Re$), the bulk Lorentz factor ($\Gamma$), the characteristic radii (photospheric radius $R_{\rm ph}$, the nozzle radius $R_{\rm 0}$, and saturation radius $R_{\rm s}$), in fact, all evolve with time. We study the thermal components in the two well-separated pulses ($P_1$ and $P_2$) by comparing their observed properties. An interesting finding in \cite{Li2023a} is that both the observational and the derived photosphere properties support the evidence that $P_1$ and $P_2$ exhibit independently pulse-wise features. $kT$, $\Re$, $\Gamma$, $R_{\rm ph}$, $R_{\rm 0}$, and $R_{\rm s}$, obtained from the thermal components in $P_1$ and $P_2$, evolve independently (see \citealt{Li2023a} for detail), suggesting the existence of two distinct thermal pulses. For instance, in both thermal components in $P_1$ and $P_2$, the blackbody kinetic temperature first decays with a PL before sharply plunging in the end. The photospheric radius $R_{\rm ph}$ and the radius at which the jet is launched to $R_{0}$ are estimated, and the saturation radius $R_{\rm s}$ consequently is given by $R_{\rm s}\equiv \Gamma R_{\rm 0}$. All these characteristic radii ($R_{\rm ph}$, $R_{0}$ and $R_{s}$) increase with time. The results are quite interesting since such distinct pulse-wise temporal features have never been clearly seen before. \cite{Ryde2004} and \cite{Ryde2005} studied the properties of thermal components with a large BATSE sample and found that $kT$ is approximately constant early on. It decays as a PL later on, while $\Re$ exhibits a PL over the whole pulse. Such general behavior is universal even for bursts with complex and heavily overlapping light curves. The thermal flux ratio ($F_{\rm BB}/F_{\rm tot}$), on the other hand, has no such clear trend. In $P_1$, it has a relatively low value ($\sim 10\%$), but increases to a relatively high value ($\sim 20\%$) quickly. However, in $P_2$, it is significantly stronger and maintains an almost constant value ($\sim 20\%$) during the entire thermal duration. Additional details regarding the observable and photospheric properties of GRB 190114C are presented in \cite{Li2023a}.

\subsubsection{Possible Afterglow Signified by $P_1$ and $P_2$} \label{subsec:PulseOnset}

GRB 190114C consists of two individual jets as indicated by the independent evolution of thermal emission in $P_1$ and $P_2$. Correspondingly, we expect the deceleration time of the first jet $t_{\rm dec}(P_{1})$ to be earlier than the deceleration time of the second jet $t_{\rm dec}(P_{2})$. Observationally speaking, at the time when the early light curve reaches a shallow decay phase of PL decay index $\sim$ -1 to -1.5 and the spectral index approximates $\sim -2$, the jet is significantly decelerated. GRB 190114C exhibits two time segments showing such behavior, the first one ($S_1$) appears after the drop of $P_2$ and the second one ($S_2$) connects to the end of $P_3$. From fitting the light curve and spectra, shown in Figure \ref{fig:Global} and Figure \ref{fig:p3_peak}, $S_{1}$ decays following a PL of $\sim -2$, and the spectral indices within its duration vary from $\sim -1.6$ to $\sim -1.9$, $S_{2}$ starts at $\sim 22$~s and continues for a later time, and the spectral indices are all $\sim -1.9$. The proposal that $S_{1}$ and $S_{2}$ are related to afterglow emission is supported by several independent studies in the literature \citep[e.g.,][]{MAGICCollaboration2019,2019A&A...626A..12R,ajello2020, Ursi2020}. We, therefore, propose that the emissions of these two segments are produced by the two decelerated jets respectively. We notice that $S_{1}$ has a steeper light-curve decay and a slightly harder spectral shape than the typical value mentioned above. This could be due to the influence of the tail of $P_{2}$. A simple extrapolation of $S_{1}$ to the time of $S_{2}$ shows the flux of $S_{2}$ is approximately three to five times higher than the extrapolated flux of $S_{1}$. This is consistent with the fact that the energy emitted in $P_{2}$ is more than three times larger than the energy emitted in $P_{1}$.

\subsubsection{Results} 

Following the physical picture in which the central engine releases two sequences of energy forming $P_{1}$ and $P_{2}$ as suggested in Section \ref{subsec:PulseWise}, and possibly two independent afterglow signatures ($S_1$ and $S_2$) corresponding to $P_1$ and $P_2$ as suggested in Section \ref{subsec:PulseOnset}, we can recalculate all the parameters based on their pulse properties (see the right panel of Figure \ref{fig:Global}). Based on our refined spectral analysis for each thermal pulse (see Section \ref{sec:SpecFit}), we obtain all the relevant observed parameters (see the upper panel of Table \ref{tab:pulse}), including the isotropic total ($E_{\gamma_1}$ and $E_{\gamma_2}$), thermal ($E_{\rm th_1}$ and $E_{\rm th_2}$), and nonthermal ($E_{\rm nth_1}$ and $E_{\rm nth_2}$) energy, the total ($F^{\rm obs}_{\gamma_1}$ and $F^{\rm obs}_{\gamma_2}$), and thermal ($F^{\rm obs}_{\rm BB_1}$ and total $F^{\rm obs}_{\rm BB_2}$) energy flux, and the BB component temperature ($kT_1$ and $kT_2$). By using these pulse-based observed parameters and individually applying the procedure proposed in \cite{Zhang2021} for each pulse, we can obtain all the fireball parameters for both $P_{1}$ and $P_{2}$. The measured parameters, including the initial dimensionless specific enthalpy ($\eta_1$ and $\eta_2$), bulk Lorentz factors at the photosphere radius ($\Gamma_{\rm ph,1}$ and $\Gamma_{\rm ph,2}$) and before fireball deceleration ($\Gamma_{0,1}$ and $\Gamma_{0,2}$), the amount of mass loading ($M_1$ and $M_2$), and GRB radiative efficiency ($\eta_{\gamma,1}$ and $\eta_{\gamma,2}$), are presented in the lower panel of Table \ref{tab:pulse}.

There are several remarks and results in order. 
(1) $E_{\gamma_i}$ is the radiated energy in $\gamma$-ray for each pulse, consisting of two parts: the MeV emission observed from \emph{Fermi}/GBM ($E^{\rm GBM}_{\gamma, \rm iso}$) and the GeV emission observed from \emph{Fermi}/LAT ($E^{\rm LAT}_{\gamma, \rm iso}$). 
(2) The amount of mass loading $M$ ($\sim$8.6 $\times$ 10$^{-4}$~$M_\odot$) derived from Section \ref{subsec:Global} is much more massive than the mass of the sum of the two thermal pulses [($M_{1}$ + $M_{2}$) $\sim$4.5 $\times$ 10$^{-4}$~$M_\odot$] presented in this Section. This is because the time intervals of thermal components in $P_{1}$ (from $t_{0}$+0.55~s to $t_{0}$+1.93~s) and $P_{2}$ (from $t_{0}$+2.45~s to $t_{0}$+5.69~s) that we used for the pulse properties are less than the entire time interval (from $t_{0}$+0~s to $t_{0}$+15~s) of prompt emission as we used for the global properties.  
(3) The derived $\eta$, $\Gamma_{\rm ph}$, and $\Gamma_{0}$ values in $P_1$ are systematically greater than those in $P_2$, and derived $M$, $E_{\rm K}$, and $E_{\rm tot}$ values in $P_1$ are systematically less than those in $P_2$, presenting a self-consistent picture in time and space as expected by the standard fireball model.
(4) A relevantly high radiative efficiency is obtained for both $P_1$ [$\eta_\gamma=(36.0\pm6.5)\%$] and $P_2$ [$\eta_\gamma=(41.1\pm1.9)\%$], suggesting that if GRBs are powered by fireballs, the efficiency can sometimes be high. More interestingly, GRB radiative efficiency obtained from $P_1$ and $P_2$ is roughly the same, which supports that the central engine of the same GRB has some common properties. 

\section{Discussion} \label{sec:Discussion}

GRB 190114C is important for the photosphere models. It has the highest value among all GRBs, in which thermal contributes $\sim 20\%$ of the prompt emission. 

Why is GRB 190114C important for the photosphere model? If only referring to the significance of the thermal emission, GRB 090902B is the most thermal dominant one to date as $\sim 70\%$ of its emission is thermal. GRB 190114C has a relatively lower, but still among the highest value in all GRBs, in which the thermal emission contributes $\sim 20\%$ of the prompt emission. The superiority of 190114C is attributed to the clear separation of its two prompt thermal pulses, while the pulses in GRB 090902B and all others are highly overlapped. Therefore, if the purpose is to investigate the activity of the central engine and the dynamics of the photospheres' evolution, GRB 190114C acts as the best example.  

Why does the high significance of thermal emission coincides with the high flux, and with the violation of the synchrotron limit? The high total flux corresponds to more photons and electrons, bringing higher opacity, consequently, higher photosphere radius, and more significant thermal emission. The intense thermal Planck spectrum adheres to the PL-like spectrum of the nonthermal emission, forming a bump-like structure, which increases the low-energy spectral index if fitting the spectrum by a PL-like function alone assuming only the synchrotron model. Indeed, the low-energy spectral index decreases when adding an additional thermal component to the fitting.

Why is the thermal percentage is almost invariable? Between $20\%$ and $40\%$ except for the very beginning, regardless of the temperature varies from $50$~keV or $200$~keV. One possible reason is the Compton upscattering moderates the ratio. This GRB occurs in one of the densest GRB environments, which has a lot of hydrogen atoms, such an environment brings high opacity for photons, and helps convert a high percentage of outflow energy to baryon kinetic energy. As a result, thermal equilibrium is maintained at a high radius, which is preferred for producing observable thermal flux. The GeV emission originates from the photons in the outflow upscatter with the dense electrons of the environment, as proposed by \cite{Beloborodov2014}. The thermal photons may contribute the most to GeV photons. This expectation of GeV origin is supported by the data, in which prompt MeV and GeV emission have a strong correlation, see Figure S8 in \cite{Wang2019}, and thermal intensity influences the MeV and GeV spectral index, see Figure 6 and Figure S8 in \cite{Wang2019}.

The initial radius $R_0$ changes significantly (See Appendix and Figure \ref{fig:deltaRph}), it increases by 2 orders of magnitude. First, different from the Lorentz factor and the photospheric radius, the evaluation of $R_0$ requires tracing the system at the observational time back to the initial state. This procedure brings a lot of uncertainty, especially influenced by the evolution of the composition and the violation of the non-dissipative assumption. Second, all the formulae are established for a single shell, without considering the energy exchange between different shells; for instance, the photosphere emission of a later shell is injected to the shell ahead. Third, the effect of equal arriving time surfaces is not cosidered. The initial time bin when the photosphere emission starts may escape from the deviation brought by the second and third reasoning. The $R_0$ derived from it carries more confident information about the black hole than the others, and its value of $\sim 10^6$~cm is consistent with the Schwarzschild radius of a typical GRB black hole of a few solar masses. $R_0$ derived from the later slices may be considered as the effective width of the shell slices (or width of photosphere), for the following reasons.

For the second possibility that the energy exchange between shells, the width of the photosphere induced by the variation of the Lorentz factor is the characteristic width in which the energy exchanges, so we can consider it as the width of a slice of shell
\begin{equation}
    \Delta R_{\rm ph} =  R_{\rm ph} \Delta\beta \simeq  \frac{R_{\rm ph} \Delta \Gamma}{\Gamma^3},
\end{equation}
and the maximum value is taken at $\Delta \Gamma = \Gamma$, then we have
\begin{equation}
    \Delta R_{\rm ph} = 10^8 {\rm cm} ~R_{\rm ph,12}~ \Gamma^{-2}_2
\end{equation}
The obtained value is in the same order of $R_0$. $\Delta R_{\rm ph}$ is approximately eight times smaller than $R_0$, see Figure \ref{fig:deltaRph}. This factor of $8$ can be eliminated by changing $Y$, the total energy versus the observed energy, and it is true that this GRB contains more energy than the isotropic energy in the prompt emission since it has strong GeV and TeV energy. $\Delta R_{\rm ph}/ R_0 \propto Y^{5/4}$, so $Y$ needs to increase five times (\citet{MAGICCollaboration2019} the computed kinetic energy of $\simeq 3 \times 10^{55}$ erg, so $Y$ can be much bigger $\sim 100$, and we can even have $\Delta \Gamma = \Gamma^{1/2}$, which is reasonable. For the fittings $\Delta R_{\rm ph} \propto \Gamma^{-3.57\pm0.14}$ and $R_{0} \propto \Gamma^{-2.97\pm0.45}$ (if $\Delta \Gamma = \Gamma^{1/2}$, the coincidence of PL indices is better), with similar evolutionary behavior.

Another possible explanation, especially for the first time bin, is due to the observational effect. The function of $R_0$ can be written as
\begin{equation}
    R_0 \propto \frac{1}{T_{\rm obs}^2} \frac{F_{\rm BB}^2}{F^{3/2}},
\end{equation}
If we forget the meaning of $R_0$, just look at the right term, which is small when the thermal flux is low and temperature is high. From the observation, we know that to have a confident thermal signature, the thermal component needs to be luminous and occupy some fraction of the total flux; otherwise, it is hidden in the data even if it exits. Also, we know that only when the total flux is high, the system is be able to produce such an observable thermal flux. Considering that the black hole suddenly ejects a lot of energy, which meets the conditions discussed above, this is the first time that there is a visible photosphere, and it is the only photosphere at that moment. Therefore, it is quite possible to observe the first thermal emission at a low photosphere radius with high temperature and low thermal ratio. As can be seen in Figure A6 in \cite{Wang2019}, the photosphere radius of the first point is the lowest and it is away from others. If we exclude the first two points of $R_0$, others are mostly in the same order of magnitude. If we cut the first and last points, the radii of Figure A6 in \cite{Wang2019} show no clear trend of decreasing or increasing.

There is a fast drop in temperature of the last slices. This is a signal that the non-dissipative assumption is not valid. The thermal energy in a shell is not only dissipating, but it also receives an energy injection from the shell behind it. The last time bin presents a shell without energy injection, and its temperature decreases even faster than the exponential cut. This decreasing behavior is much steeper than the non-dissipative prediction. 

There is a discontinuity of temperature and Lorentz factor between two pulses. If two pulses are separated, we expect the second one to almost be a repetition of the first one. If they are from a continuous process, we expect the continuities to appear in the evolution.  Although their PL trends are similar, the value of the parameters is neither repetitive nor continuous. Comparing the last points in the first pulse and the first points in the second pulse, there is a slight difference, the temperature increases, the Lorentz factor increases, and photosphere radius decreases in the first points of the second one. It is more likely that two pulses are originally independent, but still, the first one influence the second one, or the second one merges with the first cooled one.

\section{Conclusions} \label{sec:Conclusion}

In this paper, by using the uniqueness of observations of GRB 190114C and applying the new method proposed in \cite{Zhang2021}, we revisited photosphere properties and GRB radiative efficiency as first studied in \cite{Li2023a} based on an assumption that the afterglow emission starts at $\sim$ 6~s. The method described in \cite{Zhang2021} requires a GRB with a dominant thermal spectral component, a deceleration bump feature in the early afterglow light curve, and a measured redshift. One can directly dissect the GRB fireball energy budget into three components and measure their values. The observed parameters from prompt emission spectral analysis include the isotropic equivalent thermal energy $E_{\rm th,iso}$ and the isotropic equivalent non-thermal energy $E_{\rm nth,iso}$, the thermal $F^{\rm obs}_{\rm BB}$ and total $F^{\rm obs}_{\gamma}$ energy flux, and the average temperature $kT^{\rm obs}$; and from the earlier afterglow data include the deceleration time $t_{\rm dec}$ ($t_{\rm p}$). As a result, the measured parameters include the initial dimensionless specific enthalpy ($\eta$), bulk Lorentz factors at the photosphere radius ($\Gamma_{\rm ph}$), and before fireball deceleration ($\Gamma_{0}$), the amount of mass loading ($M$), the kinetic energy in the fireball $E_{\rm K}$, and GRB radiative efficiency ($\eta_{\gamma}$).

We considered two different physical pictures to revisit these properties. (1) Considering that the third pulse $P_3$ exhibits distinctly different temporal and spectral properties compared with the first two pulses ($P_1$ and $P_2$), it appears that $P_3$ is similar to the typical properties observed in afterglow emissions. We, therefore, argued that $P_3$ may originate from the RS afterglow emission by providing several pieces of additional evidence. By combining the prompt emission (by treating both thermal components observed in $P_1$ and $P_2$ as a whole) and the earlier afterglow emission ($P_3$), we derived all the parameters, and measured radiative efficiency of GRB 190114C as $\eta_{\gamma}$ = (28.3$\pm$1.4)\%. A relevantly high GRB radiative efficiency is obtained, suggesting that if GRBs are powered by fireballs, the efficiency can sometimes be high. (2) Following the physical picture in which the central engine releases two sequences of energy forming $P_1$ and $P_2$, and two possible afterglow signatures ($S_1$ and $S_2$) from $P_1$ and $P_2$, we then recalculated all the parameters based on their pulse-wise properties. Two interesting results are found. First, the derived $\eta$, $\Gamma_{\rm ph}$, and $\Gamma_{0}$ values in $P_1$ are systematically greater than those in $P_2$, and derived $M$, $E_{\rm K}$, and $E_{\rm tot}$ values in $P_1$ are systematically less than those in $P_2$, presenting a self-consistent picture in time and space as expected by the standard fireball model. Second, a relevantly high GRB radiative efficiency is obtained for both $P_1$ [$\eta_\gamma=(36.0\pm6.5)\%$] and $P_2$ [$\eta_\gamma=(41.1\pm1.9)\%$]. More interestingly, though the observed parameters are individually different (e.g., the amount of mass loading $M$), the GRB radiative efficiency obtained from $P_1$ and $P_2$ is roughly the same, which implies that the central engine of the same GRB has some common properties.

\acknowledgments

We thank the anonymous referee for the valuable comments and suggestions. We especially thank Prof. Bing Zhang for many useful discussions that greatly improved this paper and LL particularly thanks to the support from Prof. Rong-Gen Cai. This work is supported by the Natural Science Foundation of China (grant No. 11874033), the KC Wong Magna Foundation in Ningbo University, and made use of the High Energy Astrophysics Science Archive Research Center (HEASARC) Online Service at the NASA/Goddard Space Flight Center (GSFC). The computations were supported by the high performance computing center at Ningbo University.

\clearpage
\vspace{5mm}
\facilities{{\it Fermi}/GBM}
\software{
{\tt 3ML} \citep{Vianello2015}, 
{\tt matplotlib} \citep{Hunter2007}, 
{\tt NumPy} \citep{Harris2020,Walt2011}, 
{\tt SciPy} \citep{Virtanen2020}, 
{\tt $lmfit$} \citep{Newville2016}, 
{\tt astropy} \citep{AstropyCollaboration2013},
{\tt pandas} \citep{Reback2022},
{\tt seaborn} \citep{Waskom2017}}  
\bibliography{Myreferences.bib}

\clearpage
\begin{table*}
\refstepcounter{table}\label{tab:global}
\setlength{\tabcolsep}{2.5em}
\renewcommand\arraystretch{1.2}
{\bf Table~1~~ Global Properties of GRB 190114C}\\
\centering
\scalebox{0.80}{
\begin{tabular}{c|c}
\hline
\hline
Measured Parameters&$P_1$+$P_2$\\
&(From $t_{0}$+0~s to $t_{0}$+15~s)\\
\hline
Isotropic equivalent thermal energy [$E_{\rm th, iso}$]&(6.5$\pm$0.5) $\times$ 10$^{52}$~erg\\
Isotropic equivalent non-thermal energy [$E_{\rm nth, iso}$]&(2.4$\pm$0.1) $\times$ 10$^{53}$ ~erg\\
Thermal energy flux [$F^{\rm obs}_{\rm BB}$] &(1.9$\pm$0.2) $\times$ 10$^{-5}$ ~erg~cm$^{-2}$s$^{-1}$\\
Total energy flux [$F^{\rm obs}_{\gamma}$] &(1.01$\pm$0.03) $\times$ 10$^{-4}$ ~erg~cm$^{-2}s^{-1}$\\
Deceleration time [$t_{\rm dec}$]&16.4$\pm$ 0.1~s\\
Temperature [$kT^{\rm obs}$]&144$\pm$2 ~keV\\
Redshift [$z$] & 0.4254$\pm$0.0005\\
\hline
Derived Parameters\\
\hline
Dimensionless specific enthalpy [$\eta$]&708$\pm$8\\
Bulk Lorentz factor at $r_{\rm ph}$ [$\Gamma_{\rm ph}$]&666$\pm$6\\
Initial Lorentz factor [$\Gamma_{0}$]&507$\pm$5\\
Isotropic equivalent total mass [$M_{\rm iso}$]&(8.6$\pm$0.6)$\times$ 10$^{-4}$~$M_\odot$\\
Isotropic kinetic energy [$E_{\rm K, iso}$]&(7.8$\pm$0.6) $\times$ 10$^{53}$~erg\\
Isotropic total energy [$E_{\rm tot, iso}$]&(1.1$\pm$0.1)$\times$10$^{54}$~erg\\
GRB radiative efficiency [$\eta_{\gamma}$]&(28.3$\pm$1.4)~\%\\
\hline
Further Derived Parameters\\
\hline
The electron energy equipartition factor [$\epsilon_{e,-1}$]&1.36$\pm$0.03\\
The magnetic field equipartition factor [$\epsilon_{\rm B,-2}$]&0.09$\pm$0.01\\
Characteristic synchrotron frequency of afterglow emission forward shock [$\nu_{\rm m}$]&(1.9$\pm$0.2) $\times$ 10$^{17}$~Hz\\
Cooling frequency [$\nu_{\rm c}$]&(2.6$\pm$0.5) $\times$ 10$^{17}$~Hz\\
Klein-Nishina frequency [$\nu_{\rm KN}$]&(6.4$\pm$0.2)$\times$10$^{17}$~Hz\\
\hline
\end{tabular}
}
\end{table*}

\clearpage
\begin{table*}
\refstepcounter{table}\label{tab:pulse}
\setlength\tabcolsep{6pt}
\renewcommand\arraystretch{1.2}
{\bf Table~2~~ Pulse-wise properties of GRB 190114C}\\
\centering
\scalebox{0.80}{
\begin{tabular}{c|ccc}
\hline
&$P^{\rm th}_{1}$ &$P^{\rm th}_{2}$ \\
&(From $t_{0}$+0.55~s to $t_{0}$+1.93~s)&(From $t_{0}$+2.45~s to $t_{0}$+5.69~s)\\
\hline
Measured Parameters\\
\hline
Isotropic equivalent thermal energy [$E_{\rm th, iso}$]&(1.0$^{+0.7}_{-0.5}$)$\times$10$^{52}$~erg&(3.6$^{+0.6}_{-0.5}$)$\times$10$^{52}$~erg\\
Isotropic equivalent non-thermal energy [$E_{\rm nth, iso}$]&(4.8$^{+1.3}_{-1.0}$)$\times$10$^{52}$~erg&(1.3$^{+0.1}_{-0.1}$)$\times$10$^{53}$~erg\\
Thermal energy flux [$F^{\rm obs}_{\rm BB}$]&(1.5$^{+0.1}_{-0.7}$)$\times$10$^{-5}$~erg~cm$^{-2}$~s$^{-1}$&(2.3$^{+0.4}_{-0.3}$)$\times$10$^{-5}$~erg~cm$^{-2}$~s$^{-1}$\\
Total energy flux [$F^{\rm obs}_{\gamma}$]&(8.7$^{+1.6}_{-1.3}$)$\times$10$^{-5}$~erg~cm$^{-2}$~s$^{-1}$&(1.1$^{+0.1}_{-0.1}$)$\times$10$^{-4}$~erg~cm$^{-2}$~s$^{-1}$\\
Deceleration time [$t_{\rm dec}$]&$\sim$6~s&$\sim$25~s\\
Temperature [$kT^{\rm obs}$]&267$^{+22}_{-18}$~keV&145$^{+3}_{-3}$~keV\\
Redshift [$z$]& 0.4254$\pm$0.0005& 0.4254$\pm$0.0005\\
\hline
Derived Parameters\\
\hline
Dimensionless specific enthalpy [$\eta$]&$898\pm61$&$661\pm14$\\
Bulk Lorentz factor at $r_{\rm ph}$ [$\Gamma_{\rm ph}$]&$842\pm36$&$602\pm9$\\
Initial Lorentz factor [$\Gamma_{0}$]&$575\pm42$&$389\pm5$\\
Isotropic equivalent total mass [$M_{\rm iso}$]&$(1.0\pm0.3)\times10^{-4} ~M_\odot$&$(3.5\pm0.3)\times10^{-4} ~M_\odot$\\
Isotropic kinetic energy [$E_{\rm k, iso}$]&$(1.0\pm0.4)\times10^{53}$ ~erg&$(2.4\pm0.3)\times10^{53}$ ~erg\\
Isotropic total energy [$E_{\rm tot, iso}$]&$(1.6\pm0.5)\times10^{53}$~erg&$(4.1\pm0.7)\times10^{53}$ ~erg\\
$\gamma$-ray radiative efficiency [$\eta_{\gamma}$]&$36.0\pm6.5~\%$&$41.1\pm1.9~\%$\\
\hline
\end{tabular}
}
\end{table*}

\clearpage

\clearpage

\clearpage
\begin{figure*}
\centering
\includegraphics[width=0.80\hsize,clip]{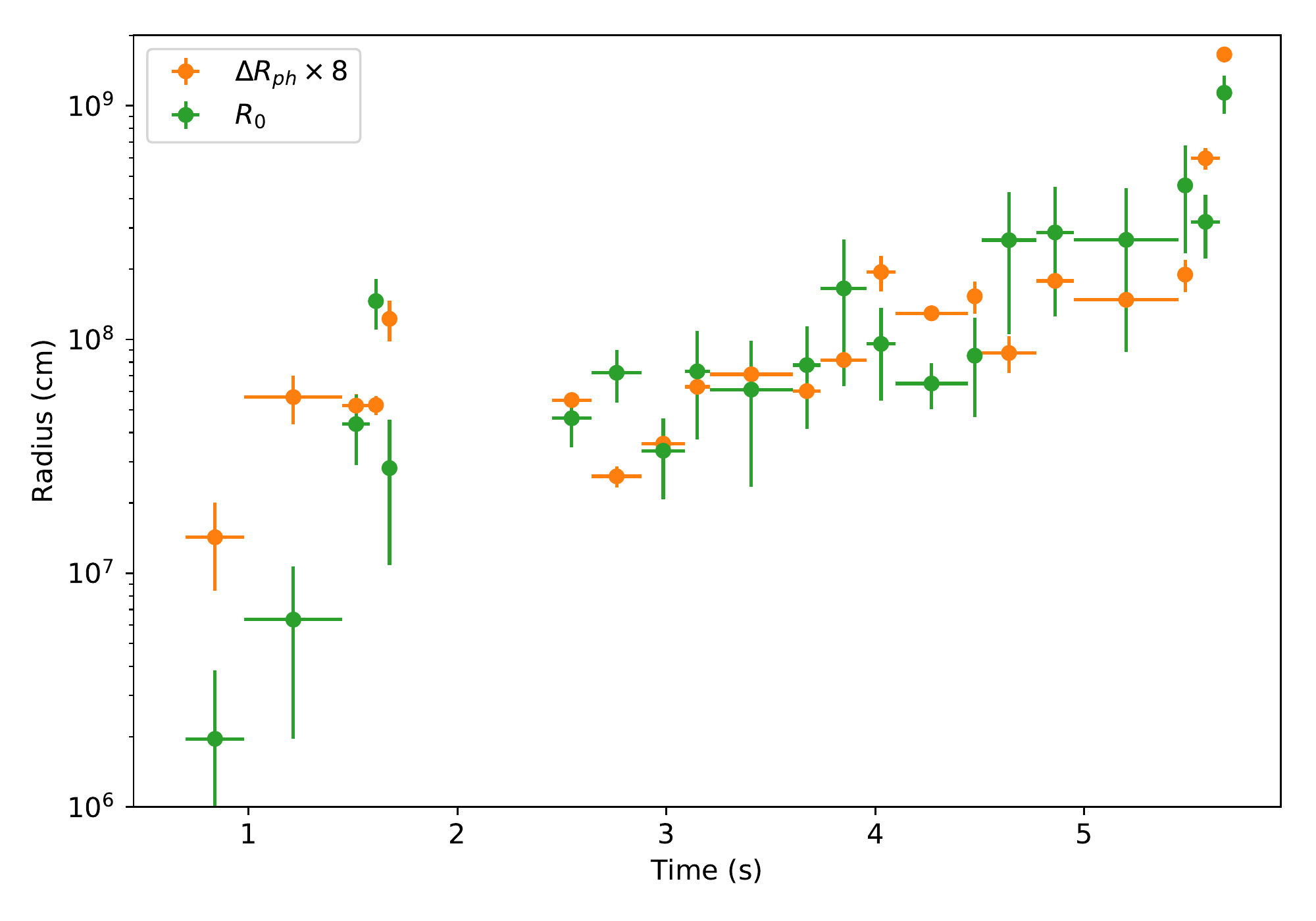} 
\caption{The temporal evolution of the nozzle radius $R_0$ and the photosphere width $\Delta R_{\rm ph}$. The photosphere width is approximately eight times smaller than $R_0$, but follows a similar trend of evolution. This factor of $8$ can be eliminated by increasing $Y$ five times, which is reasonable.}
\label{fig:deltaRph}
\end{figure*}

\clearpage
\appendix

\setcounter{figure}{0}    
\setcounter{section}{0}
\setcounter{table}{0}
\renewcommand{\thesection}{A\arabic{section}}
\renewcommand{\thefigure}{A\arabic{figure}}
\renewcommand{\thetable}{A\arabic{table}}
\renewcommand{\theequation}{A\arabic{equation}}

\subsection{Deriving the Photosphere Properties Using the Traditional Method} \label{subsec:A1}

The photosphere photons observed at a given time, corresponding to one time bin in our time-resolved analysis, are assumed to be emitted from an independent thin shell, which is given by considering the fireball optical depth falling to a unity. Therefore, the observed BB temperature $kT_{\rm obs}$, the BB flux $F_{\rm BB}$, and the total flux $F_{\rm tot}$ (thermal+non-thermal) of a given time bin determine the properties of a corresponding shell. The entire duration of photosphere emission is conjugated by the emissions from a sequence of such shells.

Within the framework of the standard fireball model~\citep{Peer2007}, for a given shell, it is generated at an initial radius
\begin{equation}
r_{0}(r_{\rm ph}>r_{\rm s})=\frac{4^{3/2} d_{\rm L}}{(1.48)^{6} \xi^{4} (1+z)^{2}}(\frac{F^{\rm obs}_{\rm BB}}{\mathbb{Y} F^{\rm obs}})^{3/2} \Re,
\end{equation}
and self-accelerates to reach a saturated Lorentz factor
\begin{equation}
\eta(\equiv\Gamma)(r_{\rm ph}>r_{\rm s})= \left[\xi (1+z)^{2} d_{\rm L} \left(\frac{\mathbb{Y} F^{\rm obs}\sigma_{\rm T}}{2 m_{\rm p} c^{3} \Re}\right)\right]^{1/4}
\label{eq:Gamma}
\end{equation}
in the coasting phase. If the photosphere radius is greater than the saturation radius, it reads 
\begin{equation}
r_{\rm ph}(>r_{\rm s})=\frac{L_{0}\sigma_{T}}{8 \pi m_{\rm p} c^{3} \Gamma^{3} },
\end{equation}
where the dimensionless parameter
\begin{equation}
\Re=\left(\frac{F_{\rm BB}}{\sigma_{\rm B} T^{4}}\right)^{1/2}=\xi \frac{(1+z)^{2}}{d_{\rm L}} \frac{r_{\rm ph}}{\Gamma}
\end{equation}
presents the effective transverse size of the photosphere. The burst luminosity $L_{0}=4\pi d^{2}_{L} \mathbb{Y} F_{\rm tot}$ is given by the observation, and $\mathbb{Y}$ is the ratio between the total fireball energy and the energy emitted in $\gamma$-rays. The numerical factor $\xi$ is of the order of unity, which can be obtained from angular integration. The luminosity distance $d_{\rm L}$ of redshift $z$ is integrated by assuming the standard FLRW metric. Other physical constants are the Thomson cross section $\sigma_{\rm T}$, the proton rest mass $m_{\rm p}$, the speed of light $c$, and the Stefan-Boltzmann constant $\sigma_{\rm B}$.

\end{document}